\documentclass[aps,prb,showpacs,twocolumn,superscriptaddress]{revtex4}
\usepackage{amsmath}
\usepackage{amssymb}
\usepackage{graphicx}

\begin{document}
\title{Common field-induced quantum critical point in high-temperature
superconductors and heavy-fermion metals}
\author{V.R. Shaginyan}\email{vrshag@thd.pnpi.spb.ru}
\affiliation{Petersburg Nuclear Physics Institute, RAS, Gatchina,
188300, Russia}
\author{M.Ya. Amusia}
\affiliation{Racah Institute of Physics, Hebrew University,
Jerusalem 91904, Israel}
\author{K.G. Popov}
\affiliation{Komi Science Center, Ural Division, RAS, Syktyvkar,
167982, Russia}
\author{V.A. Stephanovich}\email{stef@math.uni.opole.pl}
\affiliation{Opole University, Institute of Mathematics and
Informatics, Opole, 45-052, Poland}

\begin{abstract}
High-temperature superconductors (HTSC) and heavy-fermion (HF)
metals exhibit extraordinary properties. They are so unusual that
the traditional Landau paradigm of quasiparticles does not apply. It
is widely believed that utterly new concepts are required to
describe the underlying physics. There is a fundamental question:
how many concepts do we need to describe the above physical
mechanisms? This cannot be answered on purely experimental or
theoretical grounds. Rather, we have to use both of them. Recently,
in HTSC, the new and exciting measurements have been performed,
demonstrating a puzzling magnetic field induced transition from
non-Fermi liquid to Landau Fermi liquid behavior. We show, that  in
spite of very different microscopic nature of HTSC and HF metals,
the behavior of HTSC is similar to that observed in HF compounds. We
employ a theory, based on fermion condensation quantum phase
transition which is able to resolve the above puzzles.
\end{abstract}
\maketitle

The non-Fermi liquid (NFL) behavior of many classes of strongly
correlated fermion systems pose one of the tremendous challenges in
modern condensed matter physics. Many puzzling and common
experimental features of such seemingly different systems as
two-dimensional (2D) electron systems and liquid $^3$He,
heavy-fermion (HF) metals and high-temperature superconductors
(HTSC) suggest that there is a hidden fundamental law of nature,
which remains to be recognized. The key word here is quantum
criticality, taking place in quantum critical point (QCP).

Heavy fermion metals provide important examples of strongly
correlated Fermi-systems \cite{col,loh,si,sach}. The second class of
substances to test whether or not the Landau Fermi liquid (LFL)
theory is fulfilled in them, are HTSC.  In these substances, all
quantum critical points are almost inaccessible to experimental
observations since they are {\it "hidden in superconductivity"} or
more precisely, the superconductive gap opened at the Fermi level
changes the physical properties of corresponding quantum phase
transition.

There is a common wisdom that the physical properties of above
systems are related to zero temperature quantum fluctuations,
suppressing quasiparticles and thus generating their NFL properties
\cite{loh,col}, depending on their initial ground state, either
magnetic or superconductive. On the other hand, it was shown that
the electronic system of HF metals demonstrates the universal
low-temperature behavior irrespectively of their magnetic ground
state \cite{epl2}. Recently, the NFL behavior has been discovered
experimentally in 2D $^3$He, Ref. \onlinecite{he3}, and the
theoretical explanation has been given to it \cite{prlmy}, revealing
the similarity in physical properties of 2D $^3$He and HF metals. We
note here that $^3$He consists of neutral atoms interacting via van
der Waals forces, while the mass of He atom is 3 orders of magnitude
larger then that of an electron, making $\rm ^3He$ to have
drastically different microscopic properties then those of HF
metals. Therefore it is of crucial importance to check whether this
behavior can be observed in other Fermi systems like HTSC. Recently,
precise measurements on HTSC Tl$_2$Ba$_2$CuO$_{6+x}$ of magnetic
field induced transition from NFL to LFL behavior become available
\cite{pnas}. This transition takes place under the application of
magnetic field $B\geq B_{c0}$, where $B_{c0}$ is the critical field
at which the magnetic field induced CQP takes place.

Here we pay attention that to study the aforementioned transition
experimentally, the strong magnetic fields of $B\geq B_{c2}$ are
required so that earlier such investigation was technically
inaccessible. Here $B_{c2}$ is the critical magnetic field
destroying the superconductivity. We note also that an attempt to
study the aforementioned CQP experimentally had been done more then
10 years ago \cite{mack}.

In our paper, we show, that  in spite of very different microscopic
nature of HTSC and HF metals, the behavior of HTSC is similar to
that observed in HF compounds. We employ a theory, based on fermion
condensation quantum phase transition (FCQPT)
\cite{xoshag,obz1,volovik,obz} which is able to demonstrate that the
physics underlying the field-induced reentrance of LFL behavior, is
the same for both HTSC and HF metals. We demonstrate that there is
at least one quantum phase transition inside the superconducting
dome, and this transition is FCQPT. We also show that there is a
relationship between the critical fields $B_{c2}$ and  $B_{c0}$ so
that $B_{c2}\gtrsim B_{c0}$.

We have shown earlier (see, e.g. Ref. \onlinecite{obz}) that without
loss of generality, to study the above universal behavior, it is
sufficient to use the simplest possible model of a homogeneous
heavy-electron (fermion) liquid. This permits not only to better
reveal the physical nature of observed effects, but to avoid
unnecessary complications related to microscopic features (like
crystalline structure, defects and impurities etc) of specific
substances.

\begin{figure} [! ht]
\begin{center}
\vspace*{-0.5cm}
\includegraphics [width=0.45\textwidth]{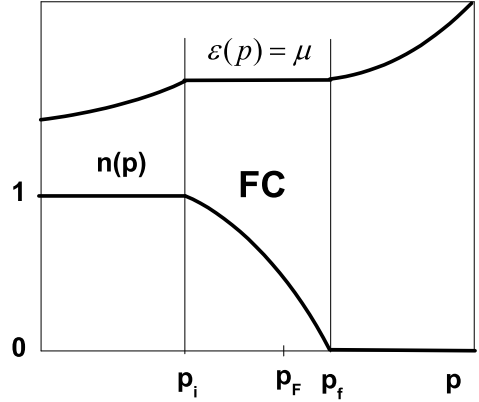}
\end{center}
\vspace*{-0.5cm} \caption{Schematic plot of the quasiparticle
occupation number $n(p)$ and spectrum $\varepsilon(p)$ in the FC
state. Function $n(p)$ obeys the relations $n(p \leq p_i) = 1$,
$n(p_i < p < p_f) < 1$ and $n(p \geq p_f) = 0$, while $\varepsilon
(p_i < p < p_f) = \mu$. Fermi momentum $p_F$ satisfies the condition
$p_i < p_F < p_f$.}\label{ff1}
\end{figure}

We consider HF liquid at $T=0$ characterized by the effective mass
$M^*$. Upon applying well-known Landau equation (see Appendix
sections for details), we can relate $M^*$ to the bare electron mass
$M$ \cite{land56, pfit} \begin{equation}\label{MM*}
\frac{M^*}{M}=\frac{1}{1-N_0F^1(x)/3}.\end{equation} Here $N_0$ is
the density of states of a free electron gas, $x =p_F^3/3\pi^2$ is a
number density, $p_F$ is Fermi momentum, and $F^1(x)$ is the
$p$-wave component of Landau interaction amplitude $F$. When at some
critical point $x=x_c$, $F^1(x)$ achieves certain threshold value,
the denominator in Eq. \eqref{MM*} tends to zero so that the
effective mass diverges at $T=0$ and the system undergoes FCQPT. The
leading term of this divergence reads
\begin{equation}\label{M1M}
\frac{M^*(x)}{M}=\alpha_1+\frac{\alpha_2}{x-x_c},
\end{equation}
where $\alpha_1$ and $\alpha_2$ are constants. At $x>x_c$ the FC
takes place. The essence of this phenomenon is that at $x>x_c$ the
effective mass \eqref{M1M} becomes negative signifying the
physically meaningless state. To avoid this state, the system
reconstructs its quasiparticle occupation number $n({\bf p})$ and
topological structure so as to minimize its ground state energy $E$
\cite{xoshag,obz1,volovik,khod,khodb}
\begin{equation}\label{FCM}
\frac{\delta E}{\delta n({\bf p})}=\mu,\end{equation} here $\mu$ is
a chemical potential. The main result of such reconstruction is that
instead of Fermi step, we have $0\leq n(p)\leq 1$ in certain range
of momenta $p_i\leq p\leq p_f$. Accordingly, in the above momenta
interval, the spectrum $\varepsilon (p)=\mu$, see Fig. \ref{ff1} for
details of its modification.

Due to above peculiarities of the $n({\bf p})$ function, FC state is
characterized by the superconducting order parameter $\kappa({\bf
 p})=\sqrt{n({\bf p})(1-n({\bf p}))}$. This means that if
the electron system with FC has pairing interaction with coupling
constant $\lambda$, it exhibits superconductivity since the
superconducting gap $\Delta\propto\lambda$ in a weak coupling limit.
This linear dependence is also a peculiarity of FC state and
substitutes well-known BCS relation
$\Delta\propto\exp{(-1/\lambda)}$, see e.g. Ref \onlinecite{bcs},
for the systems with FC \cite{xoshag,obz1,obz,amshag}.

Assume now that $\lambda$ is infinitely small. In that case, any
weak magnetic field $B$ is critical and destroys both $\kappa({\bf
p})$ and FC state. Simple energy arguments suffice to determine the
type of FC state rearrangement. On one hand, since the FC state is
destroyed, the gain in energy $\Delta E_B\propto B^2$ tends to zero
as $B\to 0 $. On the other hand, the function $n({\bf p})$,
occupying the finite interval $(p_f-p_i)$ in the momentum space,
yields a finite gain in the ground-state energy compared to that of
a normal Fermi liquid. Such a state is formed by multiply connected
Fermi spheres resembling an onion \cite{obz,zb}, see Appendix
section. In this state the system demonstrates LFL behavior, while
the effective mass strongly depends on magnetic field
\cite{obz,pog},
\begin{equation}\label{MB}
M^*(B)\propto \frac{1}{\sqrt{B-B_{c0}}}.
\end{equation}
Here $B_{c0}$ is the critical magnetic field driving corresponding
QCP towards $T=0$. In some cases, for example in HF metal
CeRu$_2$Si$_2$, $B_{c0}=0$, see e.g. Ref. \onlinecite{takah}.

\begin{figure} [! ht]
\begin{center}
\vspace*{-0.5cm}
\includegraphics [width=0.45\textwidth]{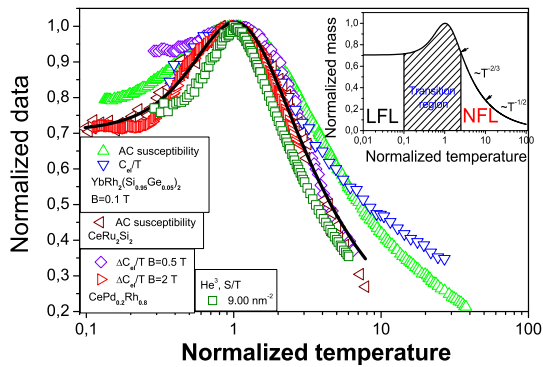}
\end{center}
\vspace*{-0.5cm} \caption{The universal behavior of $M^*_N(T_N)$,
extracted from measurements of different thermodynamic quantities,
as shown in the legend. The $AC$ susceptibility, $\chi_{AC}(T,B)$,
is taken for YbRh$_2$(Si$_{0.95}$Ge$_{0.05}$)$_2$ and CeRu$_2$Si$_2$
\cite{geg3,takah}, the heat capacity divided by temperature, $C/T$,
is taken for YbRh$_2$(Si$_{0.95}$Ge$_{0.05}$)$_2$ and
CePd$_{0.2}$Rh$_{0.8}$ \cite{cust,pikul} and entropy divided by
temperature, $S/T$, for 2D $\rm ^3He$ is taken from Ref.
\onlinecite{he3}. The solid curve gives the theoretical universal
behavior of $M^*_N$ determined by equation (\ref{fin1}) of Appendix
section. Inset shows normalized effective mass
$M^*_N(T)=M^*(T)/M^*_M$ ($M^*_M$ is the maximal value of the
effective mass at $T=T_M$) versus the normalized temperature
$T_N=T/T_M$. The hatched area outlines the transition regime.
Several regions are shown as explained in the text and Appendix
section.}\label{FD}
\end{figure}

At elevated temperatures, (see Appendix  section, eq. (\ref{fin1})),
the system transits from the LFL to NFL regime exhibiting the
low-temperature universal behavior independent of its magnetic
ground state, composition, dimensionality (2D or 3D) and even nature
of constituent Fermi particles which may be electrons or $\rm ^3He$
atoms \cite{epl2,prlmy}. To check, whether the quasiparticles are
present in the systems in the transition regime, we use the results
of measurements of heat capacity $C$, entropy $S$ and magnetic
susceptibility $\chi$. If these results can be fitted by the
well-known relations from Fermi liquid theory $C/T=\gamma_0\propto
S/T\propto \chi\propto M^*$, then quasiparticles define the system
properties in the transition regime.

As it follows from equation (\ref{fin1}), $M^*$ reaches the maximum
$M^*_M$ at some temperature $T_M$. Since there is no external
physical scales near FCQPT point, the normalization of both $M^*$
and $T$ by internal parameters $M^*_M$ and $T_M$ immediately reveals
the common physical nature of above thermodynamic functions which we
use to extract the effective mass. The normalized effective mass
extracted from measurements on the HF metals
YbRh$_2$(Si$_{0.95}$Ge$_{0.05}$)$_2$, CeRu$_2$Si$_2$,
CePd$_{1-x}$Rh$_x$, CeNi$_2$Ge$_2$ and 2D $\rm ^3He$ along with our
theoretical solid curve (also shown in the inset) is reported in
Fig. \ref{FD}. It is seen that above normalization of experimental
data yields the merging of multiple curves into single one, thus
demonstrating a universal scaling behavior \cite{epl2,prlmy,prb}. It
is also seen that the universal behavior of the effective mass given
by our theoretical curve agrees well with experimental data.

It is seen from Fig. \ref{FD} that at $T/T_M=T_N\leq 1$ the
$T$-dependence of the effective mass is weak. This means that the
$T_M$ point can be regarded as a crossover between LFL and NFL
regimes. Since magnetic field enters the Landau equation as
$\mu_BB/T$ (see Appendix  section), we have
\begin{equation}
T^*(B)=a_1+a_2B\simeq T_{M}\sim \mu_B
(B-B_{c0})\label{TB},
\end{equation}
where $T^*(B)$ is the crossover temperature, $\mu_B$ is Bohr
magneton, $a_1$ and $a_2$ are constants. In our simple model
$B_{c0}$ is taken as a parameter. The crossover temperature is not
really a phase transition. It necessarily is broad, very much
depending on the criteria for determination of the point of such a
crossover, as it is seen from the inset to Fig. \ref{FD}. As
usually, the temperature $T^*(B)$ is extracted from the field
dependence of charge transport, for example from the resistivity
$\rho(T)=\rho_0+A(B)T^2$ with $\rho_0$ is a temperature independent
part and $A(B)$ is a LFL coefficient. The crossover takes place at
temperatures where the resistance starts to deviate from the LFL
$T^2$ behavior, see e.g. Ref. \onlinecite{pnas}.

\begin{figure} [! ht]
\begin{center}
\vspace*{-0.5cm}
\includegraphics [width=0.45\textwidth]{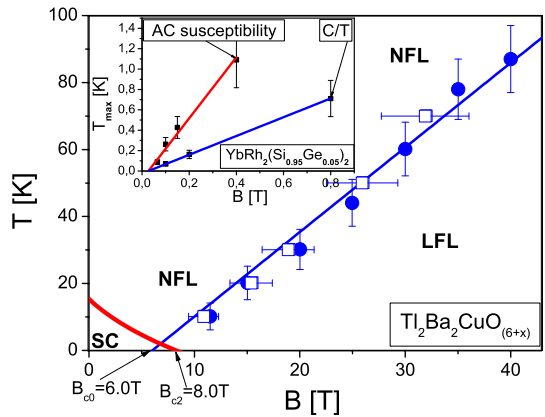}
\end{center}
\vspace*{-0.5cm} \caption{$B-T$ phase diagram of superconductor
Tl$_2$Ba$_2$CuO$_{6+x}$. The crossover (from LFL to NFL regime) line
$T^*(B)$ is given by the equation (\ref{TB}). Open squares and solid
circles are experimental values \cite{pnas}. Thick red line
represents the boundary between the superconducting and normal
phases. Arrows near the bottom left corner indicate the critical
magnetic field $B_{c2}$ destroying the superconductivity and the
critical field $B_{c0}$. Inset reports the peak temperatures $T_{\rm
max}(B)$, extracted from measurements of $C/T$ and $\chi_{AC}$ on
YbRh$_2$(Si$_{0.95}$Ge$_{0.05}$)$_2$ \cite{cust,geg3} and
approximated by straight lines (\ref{TB}). The lines intersect at
$B\simeq 0.03$ T.}\label{TMM}
\end{figure}

Let us now consider the $B-T$ phase diagram of the HTSC substance
Tl$_2$Ba$_2$CuO$_{6+x}$ shown in Fig. \ref{TMM}. The substance is a
superconductor with $T_c$ from 15 K to 93 K, being controlled by
oxygen content \cite{pnas}. In Fig. \ref{TMM} open squares and solid
circles show the experimental values of the crossover temperature
from the LFL to NFL regimes \cite{pnas}. The solid line shows our
fit (\ref{TB}) with $B_{c0}=6$ T that is in good agreement with
$B_{c0}=5.8$ T obtained from the field dependence of the charge
transport \cite{pnas}. As it is seen from Fig. \ref{TMM}, the linear
behavior agrees well with experimental data \cite{pnas}. The peak
temperatures $T_{\rm max}$ shown in the inset to Fig. \ref{TMM},
depict the maxima of $C(T)/T$ and $\chi_{AC}(T)$ measured on
YbRh$_2$(Si$_{0.95}$Ge$_{0.05}$)$_2$ \cite{cust,geg3}. As it follows
from eq. \eqref{TB}, $T_{\rm max}$ shifts to higher values with
increase of the applied magnetic field. It is seen that both
functions can be represented by straight lines intersecting at
$B\simeq 0.03$ T. This observation is in good agreement with
experiments \cite{cust,geg3}.

It is seen from Fig. \ref{TMM} that critical field $B_{c2}=8$ T
destroying the superconductivity is close to $B_{c0}=6$ T. Let us
show that this is more than a simple coincidence, and $B_{c2}\gtrsim
B_{c0}$. Indeed, at $B>B_{c0}$ and low temperatures $T<T^*(B)$, the
system is in LFL state. The superconductivity is then destroyed
since the superconducting gap is exponentially small as we have seen
above. At the same time, there is FC state at $B<B_{c0}$ and this
low-field phase has large prerequisites towards superconductivity as
in this case the gap is a linear function of the coupling constant.
We note that this is exactly the case in $\rm CeCoIn_5$ where
$B_{c0}\simeq B_{c2}\simeq 5$ T Ref. \onlinecite{pag}, while the
application of pressure makes $B_{c2}>B_{c0}$ \cite{ron}. On the
other hand, if the superconducting coupling constant is rather weak
then antiferromagnetic order wins a competition. As a result,
$B_{c2}=0$, while $B_{c0}$ can be finite as in $\rm YbRh_2Si_2$ and
$\rm{YbRh_2(Si_{0.95}Ge_{0.05})_2}$ \cite{geg3,geg}.

Upon comparing the phase diagram of $\rm CeCoIn_5$ with that of
Tl$_2$Ba$_2$CuO$_{6+x}$, it is possible to conclude that they are
similar in many respects. Further, we note that the superconducting
boundary line $B_{c2}(T)$ at lowering temperatures acquires a step,
i.e. the corresponding phase transition becomes first order
\cite{bian,epl}. This permits us to speculate that the same may be
true for Tl$_2$Ba$_2$CuO$_{6+x}$. We expect that in the NFL state
the tunneling conductivity is asymmetrical function of the applied
voltage, while it becomes symmetrical at the application of elevated
magnetic fields when Tl$_2$Ba$_2$CuO$_{6+x}$ transits to the LFL
behavior, as it predicted to be in $\rm CeCoIn_5$, Ref.
\onlinecite{pla}.

\begin{figure} [! ht]
\begin{center}
\vspace*{-0.5cm}
\includegraphics [width=0.45\textwidth]{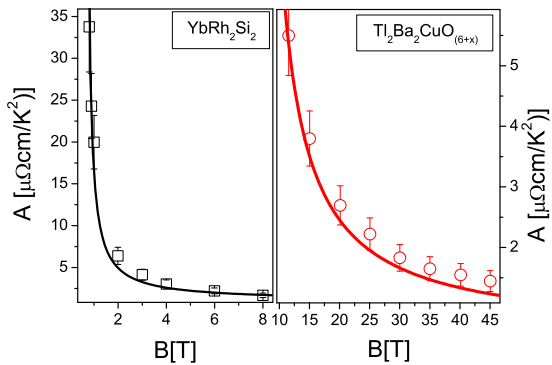}
\end{center}
\vspace*{-0.7cm} \caption{The charge transport coefficient $A(B)$ as
a function of magnetic field $B$ obtained in measurements on
YbRh$_2$Si$_2$ \cite{geg} and Tl$_2$Ba$_2$CuO$_{6+x}$ \cite{pnas}.
The different field scales are clearly seen.}\label{f2}
\end{figure}
Now we consider the field-induced reentrance of LFL behavior in
Tl$_2$Ba$_2$CuO$_{6+x}$ at $B\geq B_{c0}$. The LFL regime is
characterized by the temperature dependence of the resistivity,
$\rho(T)=\rho_0+A(B)T^2$, see also above. The $A$ coefficient, being
proportional to the quasiparticle–--quasiparticle scattering
cross-section, is found to be $A\propto (M^*(B))^2$, Ref.
\onlinecite{obz,geg}. With respect to eq. \eqref{MB}, this implies
that
\begin{equation}
A(B)\simeq
A_0+\frac{D}{B-B_{c0}},\label{HFTC}
\end{equation}
where $A_0$ and $D$ are parameters. It is pertinent to note that
Kadowaki-Woods ratio \cite{kw},  $K=A/\gamma_0^2$, is constant
within the FC theory as it follows from equations \eqref{MB} and
\eqref{HFTC}.

\begin{figure} [! ht]
\begin{center}
\vspace*{-0.5cm}
\includegraphics [width=0.45\textwidth]{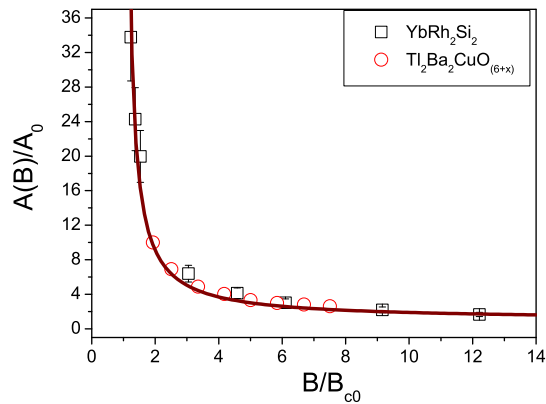}
\end{center}
\vspace*{-0.7cm} \caption{Normalized coefficient $A(B)/A_0\simeq
1+D_N/(y-1)$ as a function of normalized magnetic field $y=B/B_{c0}$
shown by squares for YbRh$_2$Si$_2$ and by circles for
Tl$_2$Ba$_2$CuO$_{6+x}$. $D_N$ is the only fitting
parameter.}\label{f3}
\end{figure}

Figure \ref{f2} reports the fit of our theoretical dependence
\eqref{HFTC} to the experimental data for two different classes of
substances: HF metal YbRh$_2$Si$_2$ (left panel) and HTSC
Tl$_2$Ba$_2$CuO$_{6+x}$ (right panel). The different scale of fields
is clearly seen as well as good coincidence with theoretical
dependence \eqref{HFTC}. This means that the physics underlying the
field-induced reentrance of LFL behavior, is the same for both
classes of substances. To further corroborate this point, we replot
both dependencies in reduced variables $A/A_0$ and $B/B_{c0}$ on
Fig. \ref{f3}. Such replotting immediately reveals the universal
nature of the behavior of these two substances - both of them are
initially in the FC state, which is being destroyed by an external
magnetic field. Since close to magnetic QCP there is no external
physical scales, the normalization by internal scales $A_0$ and
$B_{c0}$ immediately reveals the common physical nature of these
substances behavior.

In summary, it follows from our study that there is at least one
quantum phase transition inside the superconducting dome, and this
transition is FCQPT. Moreover, our consideration of above very
different strongly correlated Fermi-systems, with leading family
resemblance found between HTSC and HF compounds, shows that numerous
QCPs assumed earlier to be responsible for the NFL behavior of above
substances
can be well reduced to a single QCP related to FCQPT.\\

\noindent{\bf Appendix}\\

Consider the temperature and magnetic field dependence of the
effective mass $M^*(T,B)$ as system approaches FCQPT. Landau
equation \cite{land56} is of the form
\begin{equation}\label{sam2}
\frac{1}{M^*}=\frac{1}{M}+\int \frac{{\bf p}_F{\bf p_1}}{p_F^3}
F({\bf p_F},{\bf p}_1)\frac{\partial n(p_1,T,B)}{\partial p_1}
\frac{d{\bf p}_1}{(2\pi)^3}.
\end{equation}
The notations here are similar to those in the main text, we
suppress the spin indices for simplicity. Approximate interpolative
solution for equation \eqref{sam2} reads \cite{obz,epl2}
\begin{equation}
\frac{M^*(B,T_N,x)}{M^*_M}={M^*_N(T_N)}\approx
c_0\frac{1+c_1T_N^2}{1+c_2T_N^{8/3}}. \label{fin1}
\end{equation}
Here $M^*_N(T_N)$ is the normalized effective mass, $M^*_M$ is the
maximum value, that it reaches at $T=T_M$. Normalized temperature
$T_N=T/T_M$, $c_0=(1+c_2)/(1+c_1)$, $c_1$ and $c_2$ are fitting
parameters, parameterizing Landau amplitude. It follows from Eq.
(\ref{fin1}) that in contrast to the standard paradigm of
quasiparticles the effective mass strongly depends on temperature,
revealing three different regimes at growing temperature. At the
lowest temperatures we have the LFL regime. Then the system enters
the transition regime: $M^*_N(T_N)$ grows, reaching its maximum
$M^*_N=1$ at $T=T_M$, ($T_N=1$), with subsequent diminishing. Near
temperatures $T_N\geq 1$ the last "traces" of LFL regime disappear
and the NFL state takes place, manifesting itself in decreasing of
$M^*_N$ as $T_N^{-2/3}$ and then as $T_N^{-1/2}$. These regimes are
reported in the inset to Fig. \ref{FD}.
\begin{figure} [! ht]
\begin{center}
\vspace*{-0.5cm}
\includegraphics [width=0.45\textwidth]{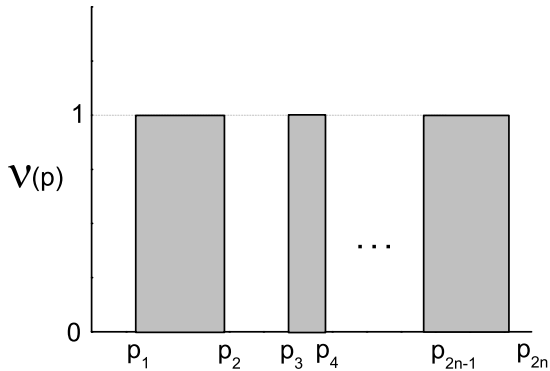}
\end{center}
\vspace*{-0.5cm} \caption{The function $\nu({\bf p})$ for the
multiply connected distribution that replaces the function $n({\bf
p})$ in the region $(p_f-p_i)$ occupied by Fermi condensate. The
momenta satisfy the inequalities $p_i<p_F<p_f$, where $p_F$ is Fermi
momentum of a normal Fermi liquid. The outer Fermi surface at
$p\simeq p_{2n}\simeq p_f$ has the shape of a Fermi step so that the
system at $T<T^*(B)$ behaves like LFL.}\label{ONI}
\end{figure}

Now we consider the action of external magnetic field on HF liquid
in FC phase. Any infinitesimal magnetic field $B \neq 0$ (better to
say, $B\geq B_{c0}$) destroys both superconductivity and FC state,
splitting it by Landau levels. The simple qualitative arguments can
be used to guess what happens to FC state in this case. On one side,
the energy gain from FC state destruction is $\Delta E_B \propto
B^2$ (see above) and tends to zero as $B \to 0$. On the other side,
$n(p)$ in the interval $p_i\leq p\leq p_f$ gives a finite energy
gain as compared to the ground state energy of a normal Fermi liquid
\cite{obz}. It turns out that the state with largest possible energy
gain is formed by a multiconnected Fermi surface, resembling an
onion so that the smooth function $n(p)$ is replaced in the interval
$p_i\leq p\leq p_f$ by the set of rectangular blocks of unit height,
formed from Heavyside step functions as reported in Fig \ref{ONI}.


\vspace*{-5mm}

\begin{thebibliography}{199}

\bibitem{col} P. Coleman and A.J. Schofield, Nature {\bf 433}, 226 (2005).

\bibitem{loh} H.v. L\"ohneysen, A. Rosch, M. Vojta, and P. W\"olfle,  Rev. Mod. Phys.
{\bf 79}, 1015 (2007).

\bibitem{si} P. Gegenwart, Q. Si, and F. Steglich, Nature Phys. {\bf 4}, 186 (2008).

\bibitem{sach} S. Sachdev, Nature Phys. {\bf 4}, 173 (2008).

\bibitem{epl2} V.R. Shaginyan, K.G. Popov, and V.A. Stephanovich, Europhys.
Lett. {\bf 79}, 47001 (2007).

\bibitem{he3} M. Neumann, J. Ny\'{e}ki, and J. Saunders, Science {\bf 317}, 1356 (2007).

\bibitem{prlmy} V.R. Shaginyan, A.Z. Msezane, K.G. Popov, and V.A. Stephanovich,
Phys. Rev. Lett. {\bf 100}, 096406 (2008).

\bibitem{pnas} T. Shibauchi
{\it et al.,} Proc. Natl. Acad. Sci. USA {\bf105}, 7120 (2008).

\bibitem{mack} A.P. Mackenzie {\it et. al.,} Phys.
Rev. B {\bf 53}, 5848 (1996).

\bibitem{xoshag} V.A. Khodel and V.R. Shaginyan, JETP Lett. {\bf 51},
553 (1990).

\bibitem{obz1} V.A. Khodel, V.R. Shaginyan, and V.V. Khodel, Phys. Rep. {\bf
249}, 1 (1994); V.A. Khodel and V.R. Shaginyan, Condensed Matter
Theories, {\bf 12}, 222 (1997).

\bibitem{volovik} G.E. Volovik, {Quantum Phase Transitions from Topology in Momentum Space},
Lect. Notes in Physics {\bf 718}, 31 (2007).

\bibitem{obz} V.R. Shaginyan, M.Ya. Amusia, and K.G. Popov,
Physics-Uspekhi {\bf 50}, 563 (2007).

\bibitem{land56} L.D. Landau, Sov. Phys. JETP {\bf 3}, 920 (1956);
E.M. Lifshitz,  and L.P. Pitaevskii, {\it Statistical Physics}, Part
2, Butterworth-Heinemann, Oxford (1999).

\bibitem{pfit} M. Pfitzner and P. W\"olfle,
Phys. Rev. B {\bf 33}, 2003 (1986).

\bibitem{khod} V.A. Khodel, JETP Lett. {\bf 86}, 832 (2007).

\bibitem{khodb} V.A. Khodel, J.W. Clark, and M.V. Zverev,
cond-mat/0806.1908.

\bibitem{bcs} J. Bardeen, L. Cooper, and J.R.
Shrieffer, Phys. Rev. {\bf 108}, 1175 (1957).

\bibitem {amshag} M.Ya. Amusia and V.R. Shaginyan, Phys.
Rev. B {\bf 63}, 224507 (2001).

\bibitem{zb} M.V. Zverev and M.J. Baldo, Journ. Phys. Condens. Matter {\bf 11}  2059 (1999).

\bibitem{pog} Yu.G. Pogorelov and V.R. Shaginyan,  JETP Lett. {\bf 76}, 532 (2002).

\bibitem{takah} D. Takahashi {\it et al.,}
Phys. Rev. B {\bf 67}, 180407 (2003).

\bibitem{geg3} P. Gegenwart {\it et.al.}, Phys. Rev. Lett. {\bf 94},
076402 (2005).

\bibitem{cust} J. Custers {\it et.al.,} Nature {\bf 424}, 524 (2003).

\bibitem{pikul} A.P. Pikul {\it et al.,}
J. Phys. Condens. Matter {\bf 18}, L535 (2006).

\bibitem{prb} J.W. Clark, V.A. Khodel, and M.V. Zverev,
Phys. Rev. B {\bf 71}, 012401 (2005).

\bibitem{pag} J. Paglione {\it et al.,} Phys. Rev. Lett. {\bf 91}, 246405
(2003).

\bibitem{ron} F. Ronning {\it et al.,} Phys. Rev. B {\bf 73}, 064519 (2006).

\bibitem{geg} P. Gegenwart {\it et al.,} Phys. Rev. Lett.
{\bf 89}, 056402 (2002).

\bibitem{bian} A. Bianchi {\it et al}., Phys. Rev. Lett. {\bf 89}, 137002 (2002).

\bibitem{epl} V.R. Shaginyan, A.Z. Msezane, V.A. Stephanovich, and
E.V. Kirichenko, Europhys. Lett. {\bf 76}, 898 (2006).

\bibitem{pla} V.R. Shaginyan and K.G. Popov, Phys.  Lett.  A {\bf 361}, 406
(2007).

\bibitem{kw} K. Kadowaki and  S.B. Woods, Solid State Comm. {\bf 58}, 507 (1986).

\end{thebibliography}
\end{document}